\documentclass[oldversion,preprint]{aa}
\usepackage{graphicx}
\usepackage{txfonts}
\begin{document}
   \title{The radio lightcurve of SN\,2008iz in M82 revealed by Urumqi observations}

   \author{N.~Marchili \inst{1} \and
           I.~Mart\'i-Vidal \inst{1} \and
           A.~Brunthaler \inst{1} \and
           T.~P.~Krichbaum \inst{1} \and
           P.~M\"uller \inst{1} \and
           X.~Liu \inst{2} \and
           H.-G.~Song \inst{2} \and
           U.~Bach \inst{1} \and
           R.~Beswick \inst{3} \and
           J.~A.~Zensus \inst{1}
          }

   \offprints{N. Marchili}

   \institute{Max-Planck-Institut f\"ur Radioastronomie, Auf dem H\"ugel 69,
              53121 Bonn, Germany\\
              \email{marchili@mpifr-bonn.mpg.de}
         \and
             Urumqi Observatory, the National Astronomical Observatories, the
             Chinese Academy of Sciences, Urumqi 830011, PR China
         \and
             Jodrell Bank Center for Astrophysics, School of Physics
             and Astronomy, Alan Turing Building, University of
             Manchester, Manchester M13 9PL
             }

   \date{Received 17 August 2009; accepted ...}

 
\abstract{We report on a set of 5\,GHz Urumqi observations of the galaxy M82, made between 
August 2005 and May 2009. From the resulting flux densities, we detect
a strong flare, starting in March or April and peaking in June
2008. We identify this flare with supernova SN\,2008iz. The time sampling of the 
radio light curve allows us to obtain information on the precursor mass-loss rate, the 
strength of the magnetic field in the radiating region, the explosion date, and the 
deceleration of the expanding shock. We also check the possible contribution of Synchrotron
Self Absorption (SSA) to the radio light curve and compare our model with other observations of 
the supernova at 22\,GHz.}


\authorrunning{N. Marchili et al.}
\titlerunning{Urumqi observations of the radio supernova SN\,2008iz}

   \maketitle
%

\section{Introduction}

The detection of a radio supernova is an uncommon event. Assuming a
typical 5\,GHz luminosity peak of $L_{\small\textrm{5\,GHz}}\sim
10^{26}$\,erg\,s$^{-1}$\,Hz$^{-1}$, the flux density of a radio
supernova at a distance of 10\,Mpc can be estimated of the order of
10\,mJy, a value which may be hard to detect with a radio antenna of small
or average size. This explains why, up to now, only about three dozen
objects are detected, while in $\sim 150$ cases only upper limits are
available (see Weiler et al. \cite{Weiler2009}).

Supernova \object{SN\,2008iz} was discovered by Brunthaler and
collaborators on April 2009 as a bright radio transient in the 
nearby galaxy \object{M82} (see Brunthaler et al.
\cite{Brunthaler2009a}, \cite{Brunthaler2009b} and \cite{Brunthaler2009c}). 
Brunthaler et al. (\cite{Brunthaler2009a}) reported on three 22\,GHz VLA 
observation epochs made between March 2008 and April 2009. These authors
found that the resulting radio light curve could be modeled using an 
exponential time decay, but could not be properly modeled using a time power-law
decay. This is an intriguing result, since the flux density decay of a supernova
is expected to follow a power-law of time (e.g. Weiler et al. \cite{Weiler2002}).

M82, at a distance of 3.5\,Mpc, is a starburst galaxy with an estimated
radio supernova rate of 1 supernova every $\sim 10 - 20$ years (e.g. Muxlow et al. 
\cite{Muxlow1994}, Fenech et al. \cite{Fenech2008}). M82 has been repeatedly observed at 5\,GHz in the
framework of the `Urumqi IDV monitoring project'. This project was
initiated in 2004 in a collaboration of the Max-Planck-Institut f\"ur
Radioastronomie with the Urumqi Observatory, as part of the National
Astronomical Observatories of the Chinese Academy of Sciences. It is
still ongoing and aims at the long term monitoring of the flux density
variability of Intraday Variable (IDV) sources (see Witzel et
al. \cite{Witzel}, Heeschen et al. \cite{Heeschen}, Krichbaum et
al. \cite{Krichbaum2002} and Kraus et al. \cite{Kraus2003}) and the
systematic search for annual modulation in type II IDV sources (see,
e.g., Rickett et al. \cite{Rickett2001}, Gab\'anyi et
al. \cite{Gabanyi}, Marchili et al. \cite{Marchili}, Marchili
\cite{MarchiliTh}). In this program, M82 (also known as 0951+699) was
regularly observed as a secondary calibrator.

The investigation of flux density variations of the order of few percent
requires that the measurements are precise to within 0.5-1\%. For
achieving such an accuracy, it is indispensable to spend a considerable amount
of the observing time on so-called calibrators -- i.e. sources whose
luminosity is constant over a time span of at least a few weeks. Due to the
stability of its emission, M82 has been included in the source lists of all
the 36 observing sessions performed in Urumqi between August 2005 and May
2009.

\begin{figure*}
\centering
\includegraphics[width=10cm]{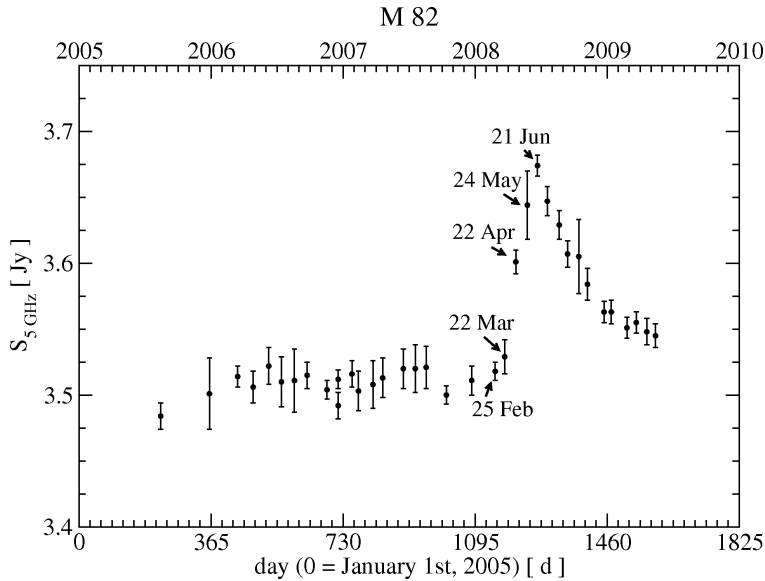}
\caption{The 5\,GHz variability curve of M82, from Urumqi observations
  performed between August 2005 and May 2009. Up to February-March 2008, the
  radio emission from the source is nearly constant. The following
  increase in the measured flux density, culminating around June 2008, has to be
  ascribed to the explosion of SN\,2008iz.}
\label{m82}
\end{figure*}

A visual inspection of the long-term variability curve of M82 revealed the
presence of a strong flare in 2008, peaking in June (see Fig.\,\ref{m82}). The
shape of the flare is consistent with a supernova explosion. The date of the
flare coincides with the explosion of SN\,2008iz (see Brunthaler
et al. \cite{Brunthaler2009c}). A deeper analysis of these data demonstrates that the
5\,GHz variability curve of M82 provided by the Urumqi telescope traces the
evolution of the radio emission from SN\,2008iz in all its phases. Thanks to the
relatively good sampling, we can estimate the parameters which characterize the
supernova explosion.

\section{Observations and data calibration}

The Nanshan radio telescope is a 25-meter parabolic antenna operated by
the Urumqi Observatory. It is located about 70\,km south of Urumqi
city at an altitude of 2029\,m above the sea level  with the
geographic longitude of 87$^\circ$E and latitude of +43$^\circ$. The
telescope is equipped with a single beam dual polarization 6\,cm
receiver (600 MHz bandwidth centered at 4.80 GHz) built by the MPIfR
and  installed at the secondary focus of the 25 m telescope.
A `Digital Backend' from the MPIfR collects data at a sampling rate of 32 ms. 
Every 32 ms the frontend setting changes, so that either a calibration signal 
of 1.7 K $T_{\rm a}$ is added to the antenna signal and/or the signal phase 
is switched off by  $180\hbox{$^\circ$ }$. This fast switching ensures a 
continuous gain control of the receiving system (for details see Sun et al. \cite{Sun}
and  references therein). The raw data are written to MBFITS format and
were analyzed at the MPIfR using a Python based software package
written by P. M\"uller. The analysis includes  a sub-scan based
combination of the 4 signal-phases, subtraction of baseline drifts and
Gaussian profile fits to the cross-scans. This analysis was done
independently for the two polarization channels (LCP/RCP) and for the
average of the two. 
Below, we give a short description of this procedure. More details can be
found in Kraus (\cite{Kraus}), Marchili et al. (\cite{Marchili}) and Marchili
(\cite{MarchiliTh}).

All the flux density measurements have been performed in cross-scan mode. Each
scan consists of 8 sub-scans in perpendicular directions over the source
position -- 4 sub-scans in elevation and 4 in azimuth.
This observing mode is particularly suitable for accurate measurements,
allowing the evaluation and correction of the pointing offsets. After removing
linear baseline drifts of the antenna response caused by the
atmosphere -- the diffuse background emission on the sky -- each
subscan was fitted by a Gaussian profile, whose amplitude provides an
estimation of the source's flux density, expressed in units of antenna
temperature (K). For the Urumqi telescope,
at 4.8\,GHz, the half-power-beam-width of the Gaussian profile is $\sim
600$\,arcsec. Given such a large beam-width, M82 -- whose angular size
is of the order of a few tens of arcseconds -- appears unresolved. This
explains why SN\,2008iz could be serendipitously observed during the Urumqi
monitoring without compromising the Gaussian profile of the sub-scans. 

After a quality-check, aiming to remove those data which do not fit standard
quality criteria, an error-weighted average of the sub-scans values provides
a single flux density measurement for each scan.
The procedure for the data calibration also corrects for gain-elevation and
gain-time effects. The former refers to the variations in the gain of the 
antenna when pointing at different elevations, due to changes in its
parabolic shape. The gain-time effect refers to unpredictable variations in the
collected flux densities, due to instabilities of the antenna-receiver system
or changing weather conditions. Both the effects can be corrected by measuring
and parameterizing the variations in the flux density of a sample of calibrators,
for the observation of which we spent about 50\% of our observing time. 

The final step of the data calibration is the conversion of the
measurements from K to Jy. The conversion factor is obtained through
the observation of sources whose flux density values at 5\,GHz are
fairly constant and well known, such as \object{NGC\,7027},
\object{3C\,286}, and \object{3C\,48}.  

\begin{table}
\caption{Urumqi data for SN\,2008iz at 5\,GHz frequency. In
  Col. 1, the observing date (at half-session); in Col. 2, the
  corresponding day (0=January the 1st, 2008).  In Col. 3 and 4,
  respectively, the flux density and the error estimations.}
\label{sn_data}
\centering
\begin{tabular}{c|c|c|c}
\hline
Observing session & Day & S$_{\textrm{sn}}$ & err \\
 (dd.mm.yyyy)     & (d) & (mJy) & (mJy) \\
\hline
25.02.2008 &   55 &  11 &  9 \\
22.03.2008 &   81 &  19 & 14 \\
22.04.2008 &  112 &  87 & 11 \\
24.05.2008 &  144 & 128 & 27 \\
21.06.2008 &  172 & 157 & 10 \\
18.07.2008 &  199 & 130 & 13 \\
20.08.2008 &  232 & 115 & 13 \\
12.09.2008 &  255 &  95 & 12 \\
13.10.2008 &  286 &  97 & 29 \\
06.11.2008 &  310 &  78 & 13 \\
22.12.2008 &  356 &  60 & 10 \\
11.01.2009 &  376 &  59 & 11 \\
23.02.2009 &  419 &  44 & 10 \\
21.03.2009 &  445 &  45 & 10 \\
19.04.2009 &  474 &  35 & 12 \\
13.05.2009 &  498 &  29 & 11 \\
\hline
\end{tabular}
\end{table}

The 5\,GHz variability curve of M82, shown in Fig.\,\ref{m82}, consists of 36
data-points, one per observing session. Each point is obtained as the mean
of the calibrated flux densities collected during the session (in
average, we have $\sim 80$ scans per session). In order to
isolate the flux emission from SN\,2008iz (S$_{\textrm{sn}}$), we need to quantify the
contribution of the rest of the galaxy and remove it. The subtraction of the
galactic emission is an easy process in the case of M82, because its 5\,GHz
flux density (S$_{\textrm{tot}}$) appears to be quite stable on a time scale of years. Between
March 2006 and December 2007 we estimated an average S$_{\textrm{tot}}$ value of
3.510\,Jy, with a standard deviation of 0.008\,Jy.

A deeper investigation of the M82 variability curve revealed the presence of a mild annual
cycle in S$_{\textrm{tot}}$ before the supernova explosion. Similar cycles have been
found also in the variability curves of other calibrators, therefore their
spurious nature is unquestionable. The origin of this systematic flux variation is still
unclear, but probably is due to residual calibration errors related to
temperature changes at the site, and to changes of the atmospheric and
ionospheric conditions. In the case of M82, this variable
component can be modeled as a sinusoidal wave with amplitude of 0.007\,Jy,
peaking in the end of June. 

The values of S$_{\textrm{sn}}$ reported in Table \,\ref{sn_data} have been calculated by
subtracting from the total flux density both the average S$_{\textrm{tot}}$ value
(3.510\,Jy) and the sinusoidal wave contribution described above.

It is worth noting that the flux density contribution from other
compact sources in M82 should not affect our results. The one known
decaying source, 41.95+575 (see Kronberg et al. \cite{Kronberg2000}),
has been shown to have a decay rate of $\sim$8.8\% per year (see
Muxlow et al. \cite{Muxlow2005}). Considering that the source, as of
July 2009, had a flux density at 5\,GHz of 10.5\,mJy (based on MERLIN
observations), its contribution to the global flux variation of M82 can
be quantified to a decline of $\sim$0.9\,mJy per year.

\section{Light curve model}

We have fitted the radio emission of supernova SN\,2008iz following 
the model described in Chevalier (\cite{Chevalier1982a}, \cite{Chevalier1982b}). 
This model relates the structure and evolution of the radio emission 
with the parameters of a hydrodynamical model of interaction between the 
supernova ejecta and the circumstellar medium
(CSM). The synchrotron radio emission is produced by the interaction of 
relativistic electrons (accelerated at the shock) from the CSM with 
magnetic fields, which are amplified in the region of expanding 
shocked circumstellar material. It is assumed that the energy density 
of the magnetic field and the number density (or energy density) of the 
electrons scale as the specific kinetic energy of the expanding shocked 
material. The radial density profile of the CSM is assumed to be 
$\propto R^{-s}$, with $s=2$. Weiler et al. (\cite{Weiler2002}) discuss 
on an analytical model of radio light curves for supernovae, based on the 
Chevalier model. Radio light curves of several supernovae have been 
satisfactorily fitted using this model (e.g., Weiler et al. \cite{Weiler2002}, 
\cite{Weiler2007}, and references therein). In our analysis, we used a 
simplified version of it. If the CSM distribution is homogeneous (i.e., there is no 
{\em clumpy} CSM) and we do not consider synchrotron 
self-absorption (SSA), the flux density of a supernova at a 
frequency $\nu$ can be modeled as

\begin{equation}
S = K_1 \left( \frac{\nu}{5~\textrm{GHz}} \right)^{\alpha} 
          \left( \frac{t-t_0}{1~\textrm{day}} \right)^{\beta} 
          e^{-\tau}
\label{EqWeiler1}
\end{equation}

\noindent where $K_1$ is a scaling factor, $\alpha$ is the spectral 
index of the emission in the optically-thin regime, $\beta$ is related 
to the flux density drop at late epochs, $t_0$ is the explosion date, 
$t$ is the epoch of observation, and $\tau$ is the opacity of the 
thermal electrons of the CSM to the radio emission. This opacity can 
be modeled as

\begin{equation}
\tau = K_2 \left( \frac{\nu}{5~\textrm{GHz}} \right)^{-2.1}
       \left( \frac{t-t_0}{1~\textrm{day}} \right)^{\delta}
\label{EqWeiler2}
\end{equation}

\noindent where $K_2$ is a scaling factor, related to the CSM density,
and $\delta$ is related to the CSM radial density profile. The exponent
$-2.1$ corresponds to the spectral dependence of free-free absorption
(FFA) by a thermal ionized gas in the radio regime.

\begin{figure*}
\centering
\includegraphics[width=11cm]{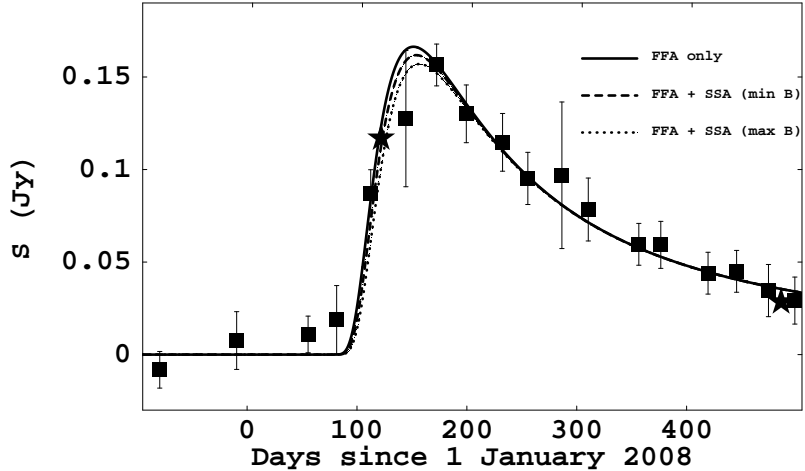}
\caption{Light curve of SN\,2008iz at 5\,GHz. Filled squares, Urumqi
data; filled stars, MERLIN data from Beswick et al. (\cite{Beswick}). The
model with only FFA is shown as a continuous line; the model with FFA
and SSA produced by the minimum (maximum) possible equipartition magnetic 
field is shown as a long-dashed (short-dashed) line.}
\label{LCFit}
\end{figure*}

On the one hand, modeling of the emission in the optically-thick 
regime (i.e., from the shock breakout to, roughly, the flux density peak) 
strongly depends on the three parameters $K_2$, $t_0$, and $\delta$. On 
the other hand, the emission in the optically-thin regime mainly depends on 
$\alpha$ and $\beta$. Our radio light curve at 5 GHz only has 2 clear 
flux density measurements in the optically-thick part of the light 
curve (see Fig. \ref{LCFit}). Therefore, it 
is numerically difficult to reliably fit our data using Eq. 
\ref{EqWeiler1}. Fortunately, $\beta$ and $\delta$ are related with the 
kinematics of the expansion, since both depend on the CSM radial density 
profile. Given that the magnetic field energy density scales as the specific 
kinetic energy of the ejecta, it is straightforward to derive that 
$\delta = \alpha - \beta - 3$. We used this relationship between $\delta$ 
and $\beta$ in the modeling of our 5 GHz data in the following way: we 
fitted these data corresponding to the optically-thin regime 
by turning off the opacity term (i.e., setting 
$\tau = 0$). Then, from the fitted value of $\beta$ and an estimate
of $\alpha$ (see next section), we derived $\delta$. Therefore, the
only remaining parameter to be fitted, when we included all the data
in the modeling, was $K_2$. With only 1 free parameter, we were able
to satisfactorily fit the optically-thick part of the radio light
curve.  

Following this approach, we arrived at slightly different results depending 
on the value of the explosion date, $t_0$, which is practically uncorrelated 
with the flux densities of the optically-thin regime and, therefore, cannot be 
well fitted using only this subset of data. However, it is possible 
to take advantage of the kinematics of the expansion to find out a 
self-consistent value of $t_0$. Let us elaborate on this: the size of 
the expanding spherical shock follows a power law of time (i.e., 
$R \propto t^m$, where $m$ is the expansion index). The expansion
index, $m$, can be related to $\beta$ and $\alpha$ as follows,
$\beta=\alpha + 3(m-1)$. From the expansion results 
based on VLBI observations recently reported in Brunthaler et al. 
(\cite{Brunthaler2009c}), we can estimate the explosion date, provided an 
estimate of $m$ is given. Therefore, we can assume an 
explosion date, $t_0$, and fit the corresponding value of $\beta$. This 
value of $\beta$ will map into a value of $m$. Then, we can derive an 
explosion date from the fitted $m$ and the VLBI results reported in 
Brunthaler et al. (\cite{Brunthaler2009c}), and look for self-consistency. 
In the next section, we report on the results obtained from this 
self-consistent fit of the radio light curve of SN\,2008iz.

\section{Results}

\subsection{Explosion Date and Expansion Rate}

The relationship between $\beta$ and $m$ depends on the spectral index
$\alpha$. Therefore, we need an estimate of $\alpha$ for our
self-consistent analysis. Brunthaler et al. (\cite{Brunthaler2009c}) report a 
spectral index of $-0.8$ for SN\,2008iz, based on VLA observations made on 27 April
2009, ranging from 1.4 to 43~GHz. However, a more careful removal of
the diffuse emission in M82 has shown a flattering of the spectrum at
1.4\,GHz, while the spectral index between 5 and 43\,GHz  is
$-1.08\pm0.08$ (Brunthaler et al. 2009, in prep.). Using 
this spectral index, together with the fit of flux densities in 
the optically-thin regime (see previous section), we find a 
self-consistent expansion index $m = 0.89 \pm 0.03$. Brunthaler et al. 
(\cite{Brunthaler2009c}) reported an explosion date on late January 
2008, assuming a constant expansion velocity between two 
VLBI epochs reported. Assuming, however, a decelerated expansion with 
$m = 0.89$, results in a delay of the explosion date by $\sim 20$ days, 
so the corrected date of the shock breakout, using a power-law decelerated 
expansion, shifts to late February 2008. If we use this new explosion date 
as an input in the fit to the data of the optically-thin part of the 
light curve, it results in $\beta = -1.43 \pm 0.05$, which maps into 
$m = 0.89\pm 0.03$. Therefore, this is a self-consistent 
result. Delaying the shock breakout to a later (earlier) date would result 
in higher (lower) values of $\beta$. This, in turn, would map into 
higher (lower) values of $m$. Since a higher (lower) $m$ corresponds 
to an earlier (later) explosion date, we can ensure that this 
self-consistency between $\beta$ and $m$ is unique. This value of $m$
is also similar to the expansion indices found in other supernovae that
could be observed with VLBI (e.g., Marcaide et al. \cite{Marcaide2009}).

Our best-fit parameters of the light curve model, obtained as described 
in the previous sections, are $K_1 = 214 \pm 4$~mJy, $\beta =-1.43 \pm 0.05$, 
$t_0$ = 18 Feb 2008 ($\pm 6$~days), $K_2 = (11.0 \pm 0.70)\times10^4$, and 
$\delta = -2.65 \pm 0.10$. The uncertainties are computed from the
diagonal elements of the post-fit covariance matrix.  

In Fig. \ref{LCFit} we show the radio light curve of SN\,2008iz at 5\,GHz,
resulting from the Urumqi measurements here reported, together with
our best-fit model (continuous line). We also show the two flux density 
measurements reported in Beswick et al. (\cite{Beswick}), based on MERLIN 
observations. These MERLIN flux densities fit remarkably well to our model. Indeed, 
Beswick et al. (\cite{Beswick}) also reported a flux density increase rate of 
$\sim$3\,mJy\,day$^{-1}$ over the two day long duration of their first
epoch observations starting 1 May 2008. Our best-fit model predicts a
flux density increase rate of $3.5 \pm 0.6$\,mJy\,day$^{-1}$ on the
same day, compatible with the rate observed by Beswick et
al. (\cite{Beswick}). This result confirms the reliability of our
light curve model.

\subsection{CSM Density and Pre-supernova Mass-loss Rate}

The parameter $K_2$ is related to the square of 
the CSM density, which is in turn related to the square of the mass-loss 
rate, $\dot M$, of the precursor star, assuming a certain stellar-wind 
velocity for the precursor (a typical value is 10~km~s$^{-1}$ for a Red 
Super Giant star). Typical values of the mass-loss rates for other 
supernovae precursor stars are $\sim 10^{-5}$~M$_{\odot}$~yr$^{-1}$. 
According to our fitting results, and using the Eq. 11 of Weiler et al. 
(\cite{Weiler2002}), the mass-loss rate of the precursor of SN\,2008iz 
would be $3.69\times10^{-5}$~M$_{\odot}$~yr$^{-1}$. To obtain this 
estimate, we have computed the ejecta velocity assuming a decelerated 
expansion with $m = 0.89$.

\subsection{Magnetic Field and Synchrotron Self-Absorption}

Since the energy density of the magnetic field and that of the relativistic electrons 
are both coupled with the specific kinetic energy of the expanding material, 
a certain equipartition of energy between particles (electrons 
and protons) and fields applies during the whole expansion. In the 
case of energy equipartition between particles and fields, it is possible 
to estimate the magnetic field in the radiating region, provided the size and 
the total luminosity of the source are known. The expression used for this 
estimate is taken from Pacholczyk (\cite{Pacholczyk1970}) (see chapter 7, 
Eq. 15):

\begin{equation}
B_{eq} = (4.5\,c_{12}\,(1+k)/\phi)^{2/7}\,R^{-6/7}\,L^{2/7}_R
\label{Beq}
\end{equation}

\noindent where $c_{12}$ depends on the spectral index, $\alpha$, and the 
minimum and maximum frequencies considered in the integration of the spectrum.
$\phi$ is the filling factor of the emitting region, $R$ is the source radius,
$L_R$ is the integrated radio luminosity, and $k$ is the ratio between the 
heavy particle energy density to the electron energy density. We do not know
the details of the particle acceleration, and the efficiency of acceleration
could depend on the particle mass. Hence, $k$ can vary from 1 to
$m_p/m_e \sim 2\times10^3$. Therefore, we are not able to compute a single 
value of equipartition magnetic field, but only a range of possibilities. 
Using the expansion information reported in Brunthaler et al. 
(\cite{Brunthaler2009c}) we can derive the source radius for the two VLBI 
epochs there reported. Then, from the flux densities recovered at those 
epochs and the source spectral index, we can estimate the range of possible 
equipartition magnetic fields at both VLBI epochs. In these estimates, we 
assume that the radio emission has a shell-like structure, with a fractional 
width of 0.3. This is the model that best describes the radio emission of 
SN\,1993J (e.g. Marcaide et al. \cite{Marcaide2009}). 

For the VLBI epoch on 3 May 2008 (63 days after explosion), we obtain a range of 
possible equipartition magnetic fields between 0.3 and 2.1 Gauss. For the 
epoch on 8 April 2009 (403 days after explosion), we obtain a range between 
0.04 and 0.31 Gauss. In these estimates, we notice that the minimum and 
maximum magnetic fields at each epoch approximately scale in time in such 
a way that the corresponding energy density is always proportional to the 
specific kinetic energy of the shock, as should be the case. All these 
values are much higher than the typical circumstellar magnetic fields 
of massive stars ($\sim$ 1~mG), thus suggesting that highly-effective 
amplification mechanisms (possibly due to Rayleigh-Taylor instabilities 
close to the contact discontinuity of the shock, see Chevalier \cite
{Chevalier1982b}) are taking place in the emitting region of SN\,2008iz.

\begin{figure*}
\centering
\includegraphics[width=11cm]{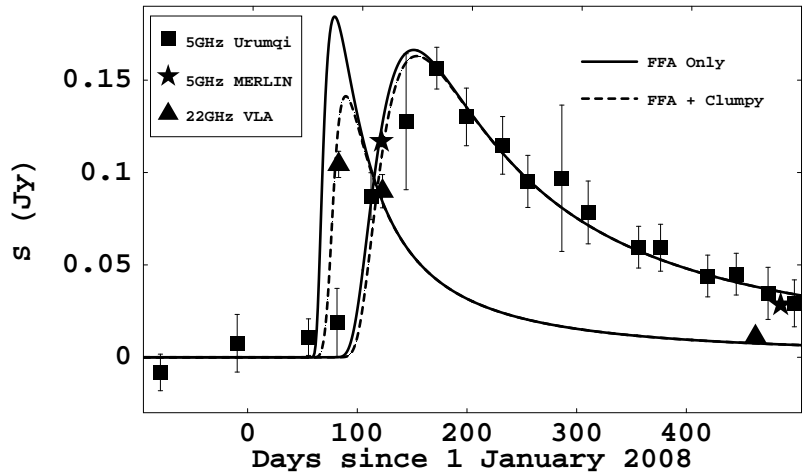}
\caption{22\,GHz flux densities taken from Brunthaler et al. 
(\cite{Brunthaler2009b}) (filled triangles), plotted along with
the 5\,GHz data shown in Fig. \ref{LCFit}. The model obtained by 
fitting the 5\,GHz data (continuous line) does not describe well the 
earliest 22\,GHz flux density. Assuming a clumpy CSM (dashed line) 
provides a better fit to this flux density.}
\label{clumpy}
\end{figure*}

The range of high magnetic fields found for SN\,2008iz translates 
into a relatively large SSA in the emitting region, thus affecting the 
behavior of the radio light curves, especially at early epochs. 
SSA effects were not considered in the model proposed by Chevalier 
(\cite{Chevalier1982b}), but were necessary to explain the radio light curves of SN\,1993J (e.g. Fransson \& 
Bj\"ornsson \cite{Fransson1998}, Weiler et al. \cite{Weiler2007},
Mart\'i-Vidal et al. in preparation). Besides the case of SN\,1993J,
SSA has not been robustly modeled in the radio light curves of any
other supernova, however tentative evidence for SSA has also been
detected in a few cases. Here, we will
compute the effects of SSA on the radio light curve of 
SN\,2008iz, and check whether the inclusion of SSA results in an 
unsatisfactory fit to these data. If that was the case, our model 
would not be self-consistent, since it would predict too high magnetic 
fields for a too low total absorption in the optically-thick regime.

To compute the SSA effects, we use the expression of the absorption
coefficient corresponding to a power-law energy distribution of electrons
(Pacholczyk \cite{Pacholczyk1970}):

\begin{equation}
\kappa_\nu = c_6\,n\,B^{1.5 - \alpha}\,
              \left(\frac{\nu}{2\,c_1}\right)^{\alpha - 2.5}
\label{Kappa}
\end{equation}

\noindent where $c_6$ depends on the spectral index, $c_1$ is a constant,
$n$ is the particle number density, and $B$ is the magnetic field. 
Assuming a constant magnetic field inside the radiating region, we can 
integrate the radiative-transfer equation easily. The flux density, 
corrected by SSA, is then

\begin{equation}
S_{SSA} = S\,\frac{1-e^{-\tau_{SSA}}}{\tau_{SSA}}
\label{SSSA}
\end{equation}

\noindent where $\tau_{SSA} = \kappa_\nu \, R_L$, being $R_L$ the 
line-of-sight depth of the source. Obviously, $R_L$ is not a constant in the
case of a spherical shell, but the use of a constant value does not affect the 
results at the precision level of our estimates. Setting $R_L$ equal to the shell 
width, in the case of a spherical optically-thick shell, gives us results which 
only deviate $\sim5\%$ from those obtained if we integrate the radiative 
transfer equation over the shell.

The only quantity that is not known in Eq. \ref{SSSA} is the electron number
density. However, we can estimate $\kappa_\nu$ from the radio luminosity of 
the source, without knowing $n$. In the optically-thin regime, the luminosity 
at a given frequency is equal to the emissivity, $\epsilon_\nu$, multiplied by 
the volume of the source. Therefore, from the luminosity and the size of the 
source, we can estimate $\epsilon_\nu$, and from the source function, 
$\epsilon_\nu / \kappa_\nu$, which is independent of $n$ (see Pacholczyk 
\cite{Pacholczyk1970}) we can determine $\kappa_\nu$. We have estimated 
the synchrotron absorption coefficient for the second VLBI epoch reported in 
Brunthaler et al. (\cite{Brunthaler2009b}), which was taken in the 
optically-thin regime at 22~GHz.

Once we have estimated $\kappa_\nu$, we can evolve $\tau_{SSA}$ by changing the 
magnetic field according to the Chevalier model, thus obtaining the 
opacity time evolution and, therefore, the SSA-corrected radio 
light curves. We show the resulting curves in Fig. \ref{LCFit}, 
for the cases of minimum and maximum possible equipartition magnetic 
fields. As it can 
be seen in the figure, the two extreme values of magnetic fields do not 
dramatically affect the quality of the fit 
(the increase of the $\chi^2$ is only 6\% for the case of minimum
magnetic field, but rises to 50\% for the case of maximum magnetic 
field). 

\subsubsection{Estimate of magnetic field from the turnover frequency}

The crossing point of the light curves at 5 and 22\,GHz, which takes 
place on day $\sim 115$ (see Fig. 3), is insensitive to slight
changes in the fitting parameters, or even to the addition of other
absorption mechanisms in the early stages of the supernova evolution,
as can be seen in Fig. 3. Therefore, we can use this point to estimate
an upper limit of the magnetic field in an alternative way than the
one followed in Sect. 4.3. Since the flux densities at 5\,GHz and
22\,GHz are equal at the  crossing point, we can use this condition to
estimate the turnover frequency at that epoch (7.81\,GHz), together
with the corresponding flux density (0.19\,mJy), assuming that SSA
dominates the absorption. Using  these estimates, together with the
supernova size interpolated at that epoch, we estimate a magnetic
field of 0.5\,G.  

Since SSA is not the dominant absorption mechanism in the light curve, 
this estimate of the magnetic field is an upper bound. The magnetic
field is inversely proportional to the turnover flux density, and such a flux 
density would be higher if there were only SSA effects in the light
curve. 

If we compare this value of the magnetic field with the equipartition 
values given in Sect. 4.3 (properly interpolated to day 115), we 
estimate a value of $k$ (see Eq. 3) of only $40 - 50$. This value of 
$k$ is also an upper bound. It indicates that the acceleration 
efficiency of the electrons by the shock might be much higher than that 
of the ions, at least until day 115. 

We note, however, that the upper bound of the magnetic field estimated 
in this section is very sensitive to the estimate of the turnover
frequency, which we have obtained by interpolation of only two spectral data
points. Any change in the estimate of the turnover frequency would translate
into changes in $k$. For instance, we could obtain equipartition between 
electrons and ions if the turnover frequency would increase to 
$\sim 15$\,GHz (assuming the same flux density), which is still between
5 and 22\,GHz. Therefore, we cannot strongly argue in favor of a low
value of $k$, based only on the crossing point of the light curves. 

\section{Comparison with other observations}

The model given by Eq. \ref{EqWeiler1} predicts the same power-law decay
of flux density at all frequencies in the optically-thin regime.
If we compare the flux densities at 22\,GHz reported in Brunthaler et
al. (\cite{Brunthaler2009b}) with the predictions of our model at 22\,GHz,
we see that the two latest data points fit remarkably well to the model. 
Indeed, according to our model, the first flux density measurement was 
made very close to the emission peak at 22\,GHz, so strong absorption effects
may be affecting the flux density of the supernova at that epoch. Indeed,
this situation could explain why Brunthaler et al. (\cite{Brunthaler2009b})
were unable to fit their 22\,GHz light curve to a time power law. The first 
flux density would not follow the expected behaviour in the optically-thin 
regime.

We notice, however, that our model predicts a too high flux density for the
first 22\,GHz epoch (a factor of $\sim$2 above the observed value). This 
could indicate that there are additional absorption mechanisms in the supernova 
that have not been considered in our modeling. Unfortunately, we have not enough 
data points in the optically-thick regime to properly model such additional 
mechanisms. Nevertheless, we can still check whether it is possible to account 
for this large discrepancy by using additional sources of absorption. Adding, 
for instance, clumpyness to the CSM with $\delta = -5$ (see Weiler et al. 
\cite{Weiler2002}) would allow to fit all the 22\,GHz flux densities 
without degrading the quality of our fit at 5\,GHz, as we show in Fig. \ref{clumpy}.
With this value of $\delta$, the clumpy medium will rapidly become homogeneous as 
the distance to the explosion center increases.

We note that this estimate of $\delta$ for a clumpy medium is not 
the result of a fit. We do not consider any clumpyness of the CSM in our best-fit 
model. Such a clumpy medium added to the model is just to show that the low quality 
fit to the earliest flux density at 22\,GHz does not necessarily indicate 
that our model is wrong. It just indicates that something may be missing in our 
modeling of the very early expansion curve for the highest frequencies.

\section{Summary}

We report on 5\,GHz Urumqi observations of the galaxy M82, in which we detected 
the radio emission from supernova SN\,2008iz as a strong flare in the galaxy 
light curve. Correcting for the galaxy contribution, we obtain the radio light
curve of SN\,2008iz. We have used the analytical model described in Weiler et al. 
(\cite{Weiler2002}) to satisfactorily fit our data. Additionally, the MERLIN 
flux density measurements reported by Beswick et al. (\cite{Beswick}), together
with the flux density rate also reported by these authors on 1 May 2008, fit 
remarkably well to our model.

From the fitted parameters of the light curve, we estimate the mass-loss rate of the 
precursor star and (using the information given in Brunthaler et al. \cite{Brunthaler2009c}, 
based on VLBI observations) we also give estimates of the explosion date, the deceleration 
of the expanding shock, and the equipartition magnetic field in the radiating region. 
We estimate a mass-loss rate of the precursor of $3.69\times10^{-5}$\,M$_{\odot}$\,yr$^{-1}$, 
similar to the typical values for other supernovae, and equipartition magnetic fields 
of the order of thousands of times larger than the typical magnetic fields around massive 
stars. This is indicative of strong amplification mechanisms in the radiating region, as 
predicted by Chevalier (\cite{Chevalier1982b}). 
Adding the contribution of SSA to the light curve does not dramatically affect the quality 
of our fit.

We compare our model predictions with the flux densities at 22\,GHz reported by Brunthaler 
et al. (\cite{Brunthaler2009b}). According to our model, the earliest flux density at 
22\,GHz may be strongly affected by absorption from the CSM. This
could have misled these authors to conclude that the flux density
decay at 22\,GHz cannot be fitted using a time power law. Indeed, the
other two flux densities at 22\,GHz fit very well to our model, in
which we apply the same power-law decay for the 5\,GHz data. Our
model, however, does not satisfactorily predict the flux density of
the earliest 22\,GHz epoch. Given that this epoch is very close to the
emission peak, we argue that other absorption effects, not considered
in our final model, may be affecting the level of emission at 22\,GHz
in the optically-thick regime. The lack of enough data at such early
epochs prevents us from a deep study in that direction.

\acknowledgements{
This paper made use of data obtained with the the 25\,m Urumqi
Observatory (UO) of the National Astronomical Observatories (NAOC) of
the Chinese Academy of Sciences (CAS). N.M. has been supported for
this research through a stipend from the International Max-Planck
Research School (IMPRS) for Radio and Infrared Astronomy at the
Universities of Bonn and Cologne. IMV is a fellow of the Alexander von
Humboldt Foundation.  
}

\end{document}